\documentclass[sigconf]{acmart}

\def\BibTeX{{\rm B\kern-.05em{\sc i\kern-.025em b}\kern-.08emT\kern-.1667em\lower.7ex\hbox{E}\kern-.125emX}}
    
\usepackage{graphicx}              
\DeclareGraphicsExtensions{.pdf,.png,.jpg,.jpeg,.PNG} 

\usepackage{microtype}                 
\PassOptionsToPackage{warn}{textcomp}  
\usepackage{textcomp}                  
\usepackage{mathptmx}                  
\usepackage{times}                     

\usepackage{graphicx}
\usepackage{amsmath}
\usepackage{amssymb}
\usepackage{multirow}
\usepackage{algorithmic}
\usepackage[ruled]{algorithm2e}
\usepackage{acronym}
\usepackage{xspace}
\usepackage[caption=false]{subfig}
\usepackage{enumitem}
\usepackage{balance}
\usepackage{setspace}
\usepackage{caption}
\newcommand{\etal}{\emph{et al.\@\xspace}}
\newcommand{\etc}{\emph{etc}.\@\xspace}
\newcommand{\ie}{\textit{i}.\textit{e}.\@\xspace}
\newcommand{\eg}{\textit{e}.\textit{g}.\@\xspace}

\acrodef{rl}[RL]{reinforcement learning}
\acrodef{irl}[IRL]{inverse reinforcement learning}
\acrodef{vr}[VR]{virtual reality}
\acrodef{ai}[AI]{artificial intelligence}
\acrodef{lfd}[LfD]{learning from demonstration}

\makeatletter
\renewcommand{\paragraph}{%
  \@startsection{paragraph}{4}%
  {\z@}{0ex \@plus 0ex \@minus 0ex}{-1em}%
  {\hskip\parindent\normalfont\normalsize\bfseries}%
}
\makeatother

\newcommand*{\origrightarrow}{}

\renewcommand*{\textrightarrow}{\fontfamily{cmr}\selectfont\origrightarrow}
\begin{document}

\copyrightyear{2019}
\acmYear{2019}
\setcopyright{acmcopyright}
\acmConference[ACM TURC 2019]{ACM Turing Celebration Conference - China (ACM TURC 2019)}{May 17--19, 2019}{Chengdu, China}
\acmBooktitle{ACM Turing Celebration Conference - China (ACM TURC 2019) (ACM TURC 2019), May 17--19, 2019, Chengdu, China}
\acmPrice{15.00}
\acmDOI{10.1145/3321408.3322633}
\acmISBN{978-1-4503-7158-2/19/05}

\title{VRGym: A Virtual Testbed for Physical and Interactive AI}

\author{Xu Xie\quad{}Hangxin Liu\quad{}Zhenliang Zhang\quad{}Yuxing Qiu\quad{}Feng Gao\quad{}Siyuan Qi\quad{}Yixin Zhu\quad{}Song-Chun Zhu}
\affiliation{%
  \institution{UCLA Center for Vision, Cognition, Learning, and Autonomy}
  \city{Los Angeles, CA 90095}}
\email{{xuxie,hx.liu,zlz,yxqiu,f.gao,syqi,yixin.zhu}@ucla.edu,sczhu@stat.ucla.edu}

\renewcommand{\shortauthors}{Xu Xie, \etal}

\begin{abstract}
We propose VRGym, a \ac{vr} testbed for realistic human-robot interaction. Different from existing toolkits and \ac{vr} environments, the VRGym emphasizes on building and training both physical and interactive agents for robotics, machine learning, and cognitive science. VRGym leverages mechanisms that can generate diverse 3D scenes with high realism through physics-based simulation. We demonstrate that VRGym is able to (i) collect human interactions and fine manipulations, (ii) accommodate various robots with a ROS bridge, (iii) support experiments for human-robot interaction, and (iv) provide toolkits for training the state-of-the-art machine learning algorithms. We hope VRGym can help to advance general-purpose robotics and machine learning agents, as well as assisting human studies in the field of cognitive science.\footnote{The code has been made publicly available at \href{https://gitlab.com/xuxie1031/vrgym}{GitLab}.}
\end{abstract}



%
\maketitle

\section{Introduction}

The past decade has witnessed a rapid development of categorical classification for objects, scenes, and actions, fueled by large datasets and benchmarks, discriminative features, and machine learning methods. Similarly, successes have also been achieved in many other domain-specific tasks, largely due to the ever-growing vast amount of labeled data and rapidly increasing computing power, combined with supervised learning methods (in particular, deep learning~\cite{hinton2006reducing}). The performance of certain tasks has reached a remarkable level, even arguably better than human in control~\cite{duan2016benchmarking,mnih2015human}, grasp~\cite{mahler2017dex,lenz2015deep}, object recognition~\cite{he2015delving}, \ac{lfd}~\cite{argall2009survey}, and playing the game of go~\cite{silver2016mastering} and poker~\cite{moravvcik2017deepstack,brown2018superhuman}.

Despite the impressive progress, these data-driven feed-forward classification methods have well-known limitations, hindering the advancement towards a more general \ac{ai} that can interact with human: (i) needing \emph{large labeled training datasets}; (ii) often \emph{task-specific} and view-dependent, which makes it difficult to generalize; (iii) lacking an \emph{explicit representation and structure} to handle large variations exhibited in and outside of the training data.

In contrast, the hallmark of machine intelligence is the capability to rapidly adapt to new tasks and ``achieve goals in a wide range of environments~\cite{legg2007universal}''. To achieve such intelligence, recent years have seen the increasing use of synthetic data and simulation platforms\footnote{See a brief review in the \href{https://www.yzhu.io/projects/turc19\_vrgym/vrgym2019turc\_supp.pdf}{supplementary}.}. Advantages include: (i) the structure of the data is efficiently encoded \emph{without the need for human labeling} as the simulation inherently comes with the ground truth; (ii) can accommodate different embodied agents (\eg, humans, humanoid robots, or turtle-bots); and (iii) benchmark \emph{generalization} in various tasks at a low cost.

\begin{figure}[t!]
    \centering
    \includegraphics[width=\linewidth]{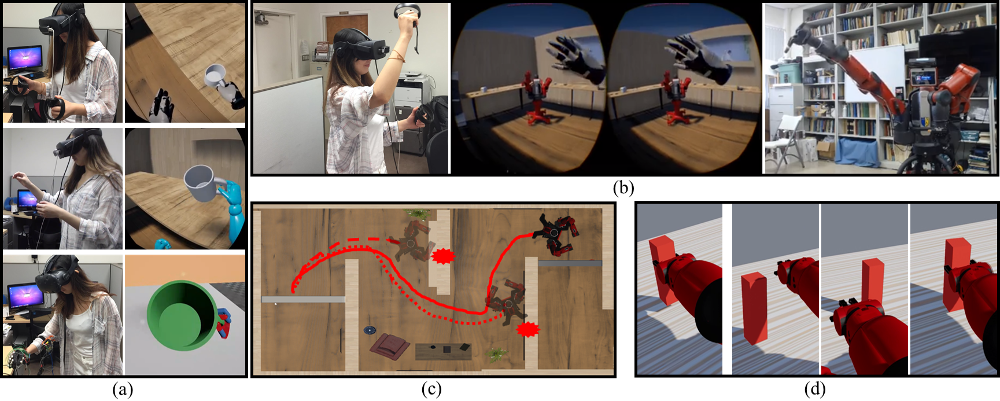}
    \caption{(a) VRGym integrates three types of input devices, providing human manipulation in an increasing resolution using Oculus Touch, LeapMotion, and a data glove, from top to bottom. (b) The VRGym-ROS bridge allows physical human/robot agent meet virtual agents inside a virtual world, providing the capability of social interactions. (c) The training of the robot navigation using \acf{rl} inside VRGym. The robot successfully navigates to the goal without collisions after about 10,000 episodes. (d) The learning of object manipulation using human demonstrations (leftmost) and \acf{irl} (right three) inside VRGym.}
    \label{fig:teaser}
\end{figure}

Empowered by the gaming industries, tremendous amount of game contents, including scenes and objects, are made available for the virtual environment. Meanwhile, more sophisticated physics-based simulation engines and rendering techniques have enabled more realistic simulations. These characteristics allow a growing number of tasks to be performed using synthetic data in simulation platforms. Furthermore, some simulation platforms also become publicly available, such as AirSim~\cite{shah2018airsim}, AI2THOR~\cite{kolve2017ai2}, Gibson~\cite{xia2018gibson}, \etc, promoting the further explorations and applications. In short, it is both the research and the engineering efforts that make it possible to achieve considerable successes in some \ac{ai} tasks and applications. 

However, prior work often lacks the human involvement, especially in high-level tasks. For instance, although some virtual platforms (\eg, OpenAI Gym~\cite{brockman2016openai} and Mujoco~\cite{todorov2012mujoco}) allow to train a virtual robot to perform many manipulation tasks, they lack a human in-the-loop, thus cannot handle critical tasks like intention prediction and social interaction. Hence, having a simulation environment where a robot can interact realistically with a human and evolve incrementally could facilitate the robotics developments.



In this paper, we propose VRGym---a virtual reality testbed, which combines \ac{vr} with virtual training for both physical and interactive \ac{ai} agents. By putting human in-the-loop, VRGym goes beyond the traditional synthetic data and simulation platforms by simulating a human-robot co-existing environment.

Specifically, VRGym tries to fill in the gap between the new advancement of \ac{vr} and the need for training virtual agents to collaborate with human. In particular, we hope to address three critical issues. First, what is the best way to reflect human embodiment in \ac{vr}; \ie, how humans can genuinely interact with robots and how the robots can perceive related data that are sufficiently close to those in real life? Second, how to take advantages of current well-developed algorithms and models? Third, to which level of unique interactions the \ac{vr} simulations can afford? To answer these questions, VRGym is designed to push the limits of current akin simulators by offering the following characteristics.

\begin{figure}[t!]
  \centering
  \includegraphics[width=\linewidth]{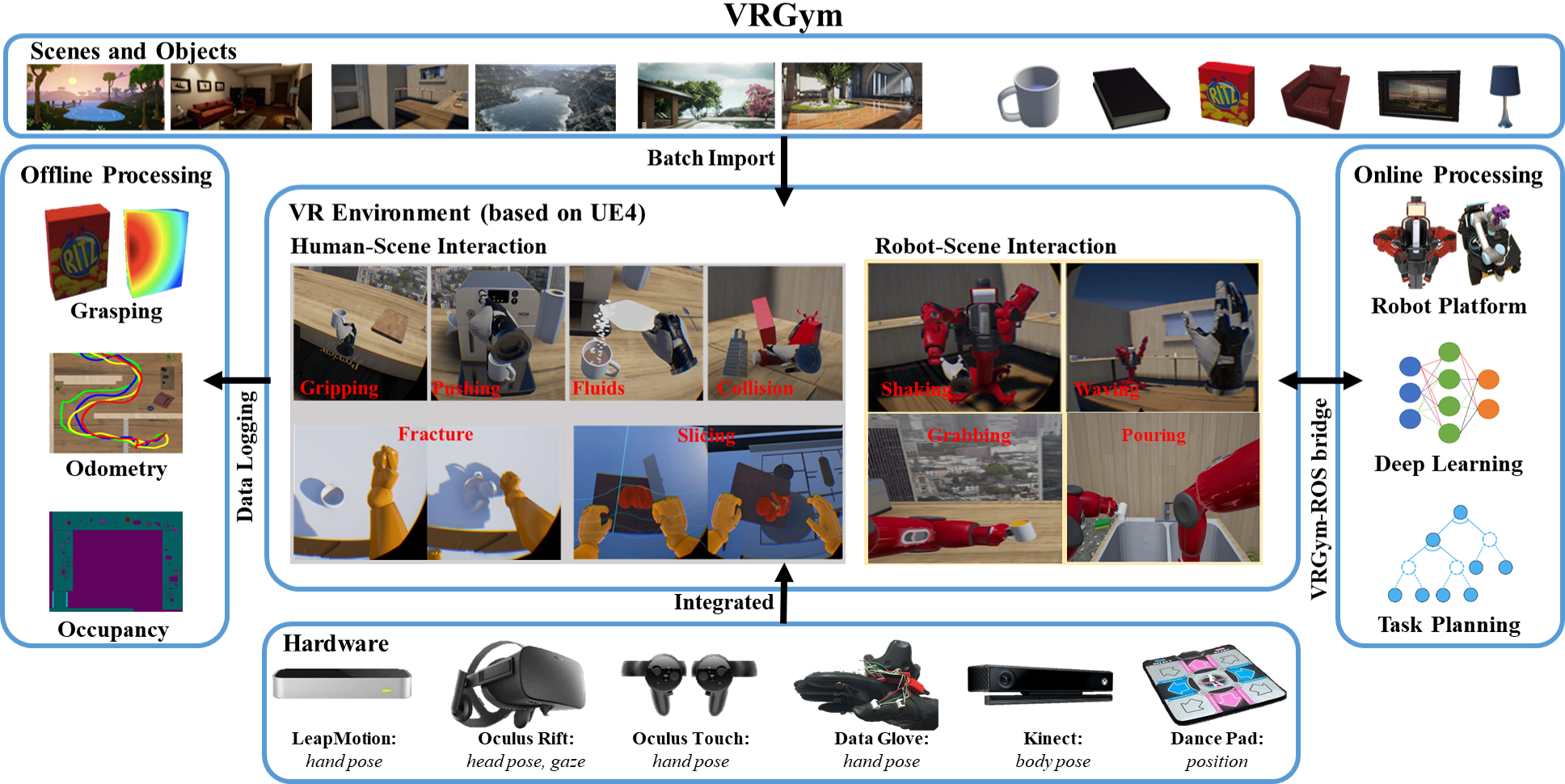}
  \caption{System architecture of VRGym, consist of three major components: (i) Hardware modules for human data input. (ii) Scene modules batch import various category of scenes as well as diverse objects, derived from different resources such as 3D modeling tools, scanned models, and automatically generated synthetic data. (iii) \ac{vr} environment serves as an ideal testbed, where both a human and a robot can perform diverse tasks. The inherent physics-simulation engine enables realistic human-scene interactions and robot-scene interactions.}
  \label{fig:system}
\end{figure}

\paragraph{Fine-grained human embodiment representation}

Adding a real human in the simulation is not a trivial task. Most of the current simulation platforms only support either scripted or limited remote-controlled human models. In VRGym, we integrate a multi-sensor setup as alternatives to traditional \ac{vr} input devices. Our setup is capable of providing a whole-body sensing and reflecting the measured data on a detailed human avatar. As a result, the simulation can account for both body and hand poses during interactions. \autoref{fig:teaser}a shows different resolutions of manipulations in VRGym.

\paragraph{High compatibility with existing robotics systems and algorithms}

In VRGym, we build an efficient bi-directional communication interface with the Robot Operating System (ROS). \autoref{fig:teaser}b depicts an example of how a person interacts with a robot in VRGym, supported by the VRGym-ROS bridge. As a result, all ROS-compatible resources can be used in VRGym with little effort, which allows easy setups, training, evaluations, and benchmark.

\paragraph{Multiple levels of interactions}

By providing the fine-grained human embodiment representation and the ROS integration, various interactions between humans and autonomous agents are made possible in different resolutions. VRGym supports interactions as simple as only providing visual/perception information and as sophisticated as learning complex robot grasping from human demonstrations. \autoref{fig:teaser}c shows how an agent obtains a navigation policy using \ac{rl}, and \autoref{fig:teaser}d shows learning a grasp policy using \ac{irl}.

VRGym makes the three contributions:
\begin{itemize}[leftmargin=*,noitemsep,nolistsep]
    \item A comprehensive simulation platform that integrates UE4 built-in functions, \eg, scene, physics-based simulation, rendering, basic human inputs, with customized developments, aiming to facilitate a variety of \ac{ai} researches.
    \item A multi-sensor hardware and software setup that allows the whole body sensing and reflects human subjects to virtual embodiments with great details. The generated data can be seamlessly logged for online and offline training purposes.
    \item VRGym-ROS bridge enables a bi-directional data communication. Through this interface, \ac{ai} researchers can take advantages of the existing robotics models and algorithms. Similarly, robotics researchers can utilize more sophisticated physics-based simulation.
\end{itemize}


\section{VRGym System Architecture}\label{sec:system}

\autoref{fig:system} illustrates the system architecture of the VRGym. VRGym offers a variety of realistic scenes and tasks for both humans and robots, and provide automatic logging of the data during agents performing tasks. This capability is provided by the integration of three main modules: (i) scene module which renders user-specified 3D scenes and objects, (ii) \ac{vr} environment based on UE4 with physics-simulation engine, introducing various physical properties that enrich tasks and data, and (iii) \ac{vr} hardware module that imports a human agent's state and command to the VRGym. We now further elaborate each module in the following subsections.

\subsection{Scene Module}


Scenes and objects are the building blocks for a simulation environment. In order to increase the variety of environments for VRGym, we develop several pipelines to import or create scenes into VRGym based on the users' specifications. The scene module enriches static environments for VRGym. Note that the ground truth of RGB image, depth image, surface normal, and object label come automatically with the scene module in real-time, enabling the training for machine learning models and robotics applications.

Specifically, VRGym can directly import the entire 3D scenes provided in large open-source datasets, either collected from the web~\cite{song2017semantic,chang2017matterport3d} or automatically generated from a given set of objects~\cite{yu2011make,qi2018human,jiang2018configurable} (see top of \autoref{fig:system}). Additionally, VRGym also supports manually constructed scenes (see \autoref{fig:task_rich}) for more specific tasks, where neither the open-source scene dataset or the automatically generated scenes could satisfy such constraints.

Similarly, individual objects can be imported to VRGym from mesh files, which can be obtained from open-source CAD datasets (\eg,~\cite{chang2015shapenet,calli2015benchmarking}). Customized or complex objects can be manually created or scanned using a RGB-D sensor to import to VRGym for specific tasks. After the import, users can further adjust static meshes, textures, materials, and collision boundaries of the objects.

\subsection{Real-time Physics-based Simulation}

\begin{figure}[t!]
    \centering
    \includegraphics[width=\linewidth]{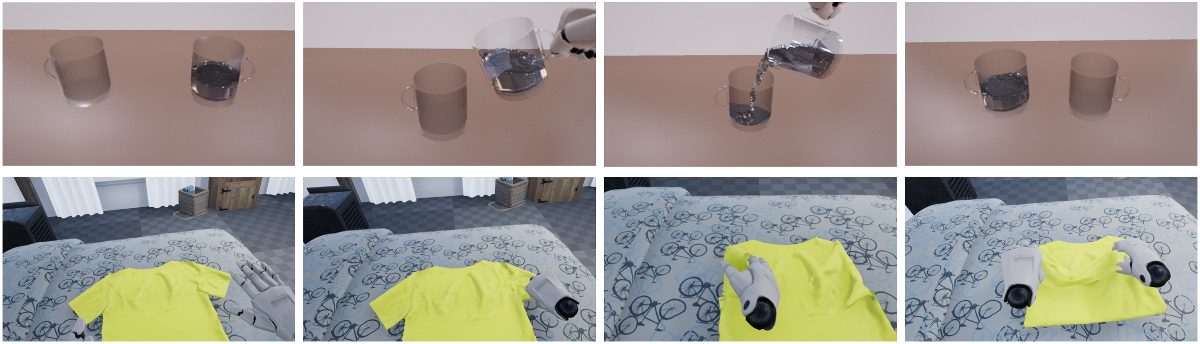}
    \caption{Examples of various physics-based simulation for diverse tasks in VRGym beyond merely rigid-body simulation in other 3D virtual environments. (Top) Pouring water. (Bottom) Folding clothes.}
    \label{fig:simulation}
\end{figure}

We choose UE4 as the simulation engine for VRGym for its advanced real-time physics-based simulation. Unlike previous 3D virtual environments that mostly focus on rigid body simulation or symbolic-level event simulation, VRGym integrates the advanced simulation provided by UE4 to enable a large set of various simulations, including rigid body, soft body, collision, fluid, cloth, slicing, and fracture. Some examples are shown in \autoref{fig:simulation} and the center of \autoref{fig:system}. As a result, subtle object state or fluent~\cite{newton1736method} changes due to the virtual agent's actions are realistic and diversify. Integrating with such sophisticated physics-based simulations, VRGym not only increases the task complexity and improves the visual experience of human agents, but also affords more complicated task simulations for both virtual and physical robots.

\subsection{Human Embodiment in VRGym}

Compared to other similar 3D virtual environments, VRGym has another distinct feature; \ie, introducing the capability to represent the physical human agent's embodiment in real-time as an avatar in the virtual environment. To reflect human movements and manipulations accurately, the physical human agent is tracked in real-time, resulting in a humanoid mesh that can deform accordingly based on the underlying tracked body skeleton and the hand poses.

Specifically, the setup includes: (i) A Kinect One RGB-D sensor to map human skeleton to the avatar in real-time through a customized-built Kinect plugin developed in UE4, (ii) an Oculus headset to record the head pose, (iii) a dance pad to navigate the avatar inside a large virtual world, and (iv) three types of input devices that provide manipulation information in different resolutions. Compared to other platforms, VRGym emphasizes the capability for users to interact with virtual environments. Depending on the needs, the user can use one of the three input devices for manipulation:
\begin{itemize}[leftmargin=*]
    \item Oculus Touch Controller offers an attachment-based approach; \ie, the virtual object will automatically attach to the virtual controller/hand once the user triggers the grasp event. It enables a firm-grip manipulation, providing a firm but the least realistic grasp during the human-object interaction. Such manipulation is effective in the event-level tasks where the fine-grained hand pose is not required; \eg, pick and place.
    \vspace{6pt}
    \item The commercial hand pose sensing products (\eg, LeapMotion) provide the vision-based gesture recognition. It is a low-cost and off-the-shelf solution that can be easily set up by mounting the sensor on the head-mounted display. However, it is difficult to have a firm grasp due to occlusions and sensor noises. Note that the hand tracking will fail if the hand is not within the view.
    \vspace{6pt}
    \item An open-sourced glove-based device~\cite{liu2019high} is also compatible with VRGym to provide the finest-grained manipulation. It requires a Vive Tracker to provide global positioning of the hand, and an IMU network in the glove to measure the rotation of each phalanx and calculate the hand poses using forward kinematics. Although glove-based devices are costly compared to other alternatives, they allow reliable hand pose sensing, which is vital for the tasks with detailed, complex and subtle hand manipulations.
\end{itemize}

\begin{figure}[t!]
  \centering
  \includegraphics[width=\linewidth]{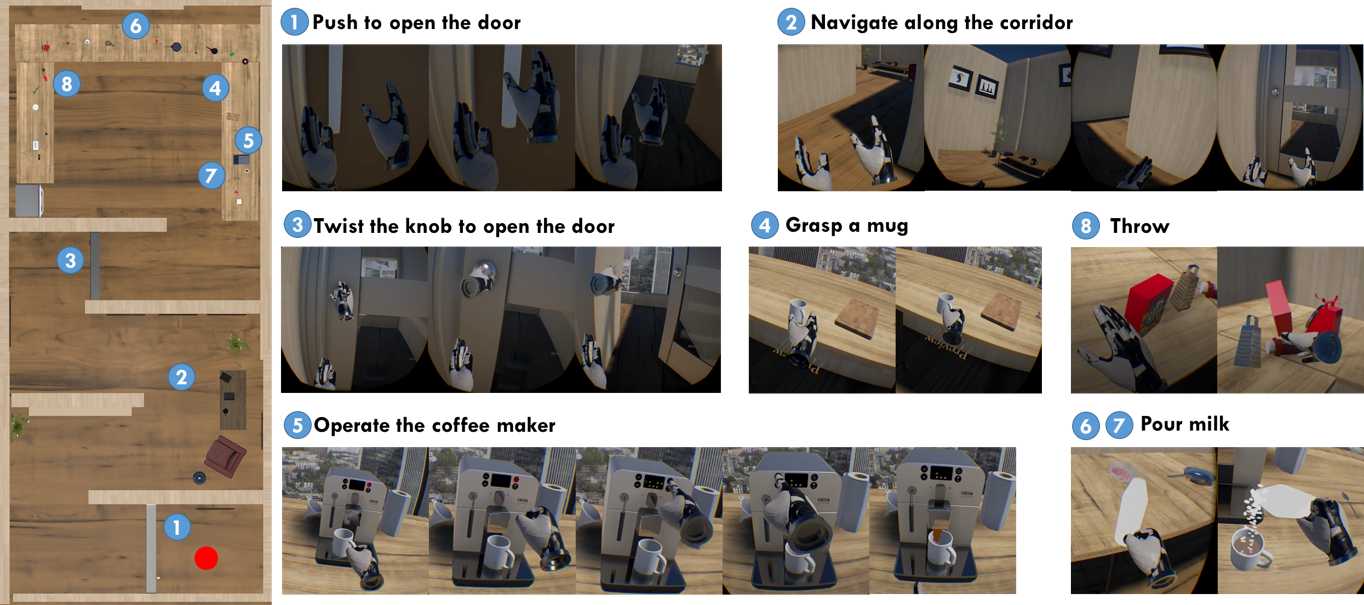}
  \caption{A human agent performs a series of actions in a virtual scene using Oculus Touch controllers. (Left) Action sequence from a top view of a virtual indoor environment. (Right) Sequences of the performed actions. Specifically, the human agent starts at the red dot as shown in the left, (1) pushes a door, (2) navigates along the hall, (3) twists a door to enter the kitchen, and (4)-(7) makes a cup of coffee. This process involves (i) large movements using the human embodiment provided in VRGym (navigating along the hallway), (ii) complex operations (operating the coffee maker), (iii) fine-grained manipulations (twisting the doorknob), and (iv) physics-rich controls (pouring milk).}
  \label{fig:task_rich}
\end{figure}

\section{Software Interface Design}\label{sec:interface}

VRGym has two major software interfaces developed to enable training and benchmarking both physical and interactive \ac{ai} agents. The first interface is the human data logging system that builds on top of the hardware setups to collect the data generated during the interactions between the avatar and the environment. Another interface, a VRGym-ROS bridge, is introduced to allow seamlessly import of robot models and robotics algorithms from ROS. The collected data together with the VRGym-ROS bridge could be used for a variety of \ac{ai} applications; see examples in \autoref{sec:experiment}.

To demonstrate the functions of these two interfaces, we consider a task-rich environment built for the VRGym. \autoref{fig:task_rich} depicts an environment in VRGym that provides semantically-diverse tasks to the agent. Note that although such environment could be constructed in the real world to perform the demonstrated tasks, sensing and logging the detailed data generated during the interactions between the agent and objects would be extremely difficult in practice.

In such a typical virtual environment in VRGym, an agent (a human as an avatar or a virtual robot) is initially placed on the starting point, indicated as the red dot in \autoref{fig:task_rich}. The final goal for the agent is to reach the kitchen located at the far-end, and accomplish several sub-tasks. At the beginning, the agent has to push to open the first door and navigate along the corridor, requiring \emph{large movements}. Then the agent must go through another door to enter into the kitchen, and the only solution is to twist the doorknob using complex \emph{manipulations}. Inside the kitchen, the agent is required to make a cup of coffee with milk, which needs to grasp and move a mug, operate the coffee maker by pushing several buttons in a certain order. The entire procedure requires the \emph{task planning} empowered by the \emph{physics-based simulation}. 

\subsection{Human Data Logging}

When a user performs a task, data generated by the interactions between an avatar and the environment can be directly logged with ground-truth labels in VRGym. In this section, we showcase two scenarios where the data is logged and used in other applications.

\begin{figure}[t!]
    \centering
    \includegraphics[width=\linewidth]{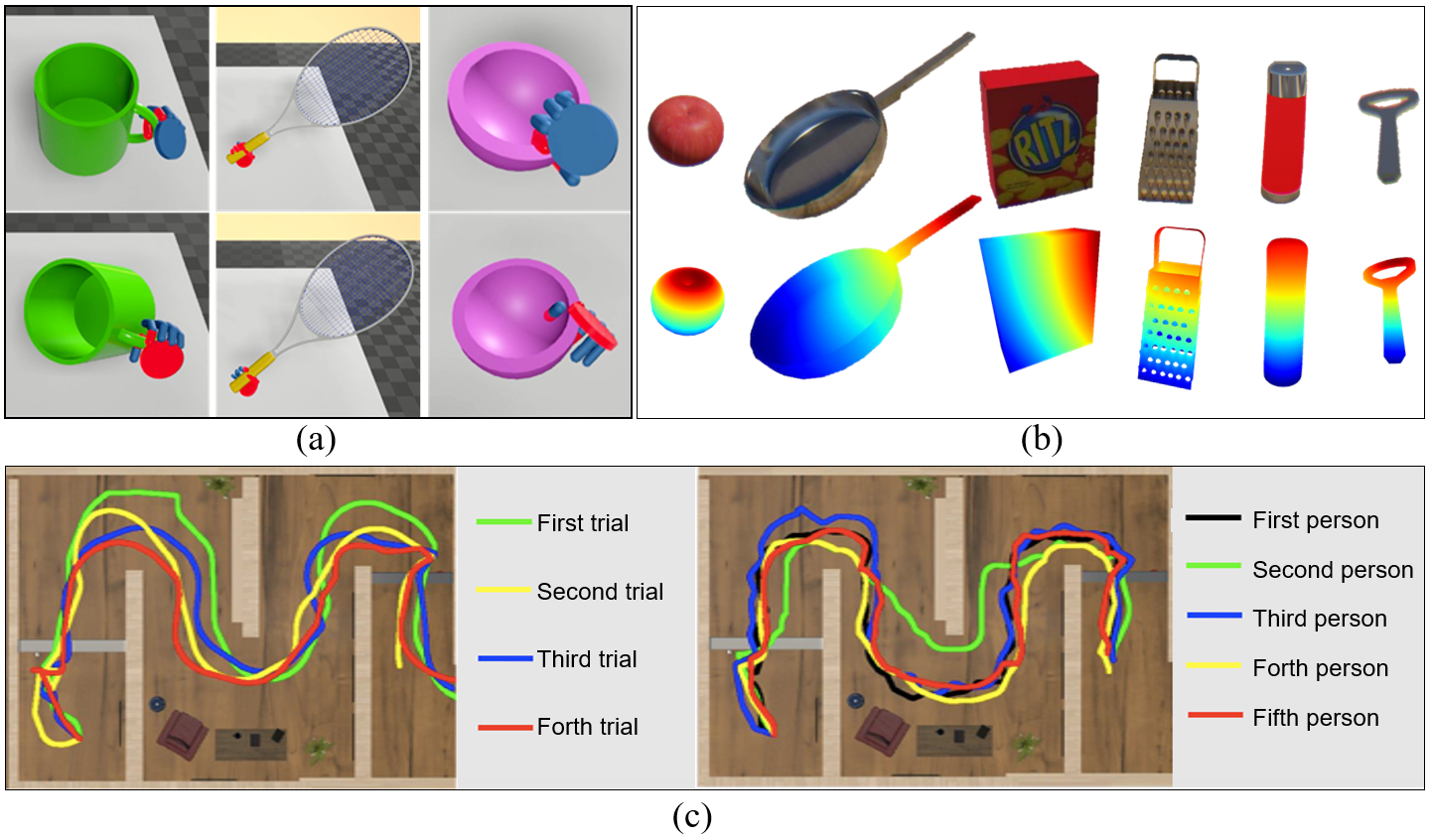}
    \caption{(a) Grasp a mug, a tennis racket, and a bowl. The red area indicates the contact force between the virtual hands and the object. (b) Visualization of the collected human grasp data. Top: a set of 3D objects. Bottom: the average grasp heat map generated by multiple subjects. (c) Visualization of footprint from different subjects.}
    \label{fig:logging}
\end{figure}

\paragraph{Grasping}

Finer-grained manipulation is made feasible in VRGym using a glove-based device~\cite{liu2019high}; see \autoref{fig:logging}a for some results. By collecting a set of subjects' grasp data on a variety of objects, we can merge all the collected grasp data to form heat maps on different objects to visualize the likelihood of grasp points on man-made objects. Specifically, the grasp data shown in \autoref{fig:logging}b is the averaging data of heat maps collected from 10 human subjects, where the hotter the area is, the denser the grasp points are, and the more likely a human agent would grasp around that area. 


\paragraph{Footprints}

VRGym provides the function to log an agent's footprints or the odometry data. \autoref{fig:logging}c shows the recorded odometry data from 5 human subjects who have limited \ac{vr} experience. Each of the participants navigates from the starting point to the kitchen room along the corridor using Oculus Touch controllers.


\subsection{ROS Interface}

The VRGym is compatible with the popular ROS framework through a customized VRGym-ROS communication bridge. This bridge allows the off-the-shelf ROS robot models to communicate with the simulations and human agents in VRGym with minimal efforts; \eg, the diverse scenes rendered in VRGym can also be exported to the Gazebo simulator, which is highly compatible with ROS.

\paragraph{Implementation}

We develop a ROS interface, VRGym-ROS bridge, based on the TCP/IP protocol in order to enable VRGym to communicate with the existing popular robotics platforms. Through this interface, robot body parts can be easily imported to \ac{vr} environments as mesh files and control signals, and a data stream can be seamlessly transferred between the VRGym and the robot platforms using ROS to communicate with either physical or virtual robots. We organize all data types (\ie, ROS topics) in a unified JSON format and construct JSON parsers in both VRGym and ROS to further improve the compatibility. Each port in the protocol supports a stream of data, making it possible to present multiple agents from ROS into the VRGym. With the VRGym-ROS bridge, we present two examples of training and evaluating human robot interactions (HRI) inside VRGym in \autoref{sec:experiment}, which incorporates direct human reactions and involvements. Such capability is largely missing in the current robotics simulators such as Gazebo or V-Rep. The benchmark in \autoref{sec:experiment} is also supported by this VRGym-ROS bridge.

\begin{figure}[t!]
    \captionsetup[subfloat]{farskip=0pt}
    \centering
    \subfloat[]{\includegraphics[height=0.285\linewidth]{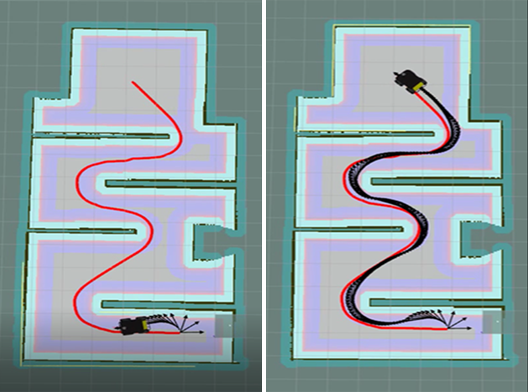}\label{fig:gazebo_navigation}} \hfill
    \subfloat[]{\includegraphics[height=0.285\linewidth]{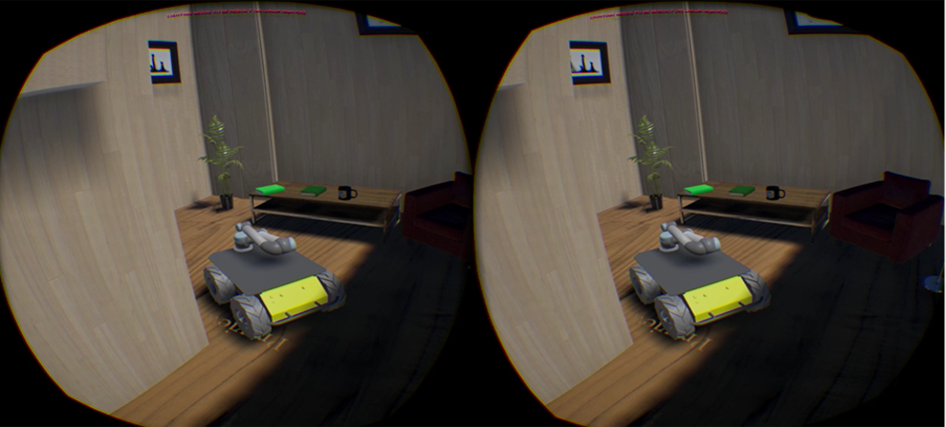}\label{fig:ur5}} \hfill
    \caption{VRGym-ROS bridge. (a) The robot navigation in the scene imported into the Gazebo, exported from the VRGym. The red curve indicates the path planned by the robot's global planner. The black curve is the actual trajectory executed by the robot. (b) A Husky-UR5 robot is imported into VRGym from ROS to guide the way and open the door for a human agent.}
\end{figure}

\paragraph{Evaluation}

We evaluate the VRGym-ROS bridge on a navigation task (see \autoref{fig:task_rich}) using a Clearpath Husky robot. This navigation task is performed in VRGym, whereas the robot model is imported from ROS, making it possible to evaluate a number of SLAM algorithms and path planning approaches. In \autoref{fig:gazebo_navigation}, the mapping result is obtained using the conventional GMapping package in ROS. The red curve indicates the planned path, whereas the black curve is the actual odometry of the Husky robot. \autoref{fig:ur5} shows the user's view when the robot is moving. This VRGym-ROS bridge fills in the gap between the diverse scenes in VRGym and the existing fine-tuned algorithms provided in ROS.

\paragraph{Communication Bandwidth}

To evaluate the reliability and efficiency of the VRGym-ROS bridge, we conduct an experiment by sending packages with the size of 512Kb\footnote{See a detailed evaluation in \href{https://www.yzhu.io/projects/turc19\_vrgym/vrgym2019turc\_supp.pdf}{supplementary}.}.

\section{Experiments}\label{sec:experiment}

In this section, we demonstrate the performance and capability of the VRGym from four different perspectives\footnote{See a video demo at \href{https://vimeo.com/327629749}{Vimeo}.}. Two human robot interaction (HRI) applications are conducted, including a human intention prediction task and a social interaction task. Like other testbeds, we also benchmark the performance popular machine learning algorithms (\eg, reinforcement learning and \ac{irl}) in the VRGym.

\subsection{Experiment 1: Intention Prediction}\label{sec:intention}

Predicting human intention is difficult when training on a physical robot since this task has very small error tolerance; wrong predictions may endanger both the human and the robot. It is particularly interesting to study human intention prediction in VRGym, since this problem involves complicated inference process that many types of data can be useful: human trajectories, human poses, object positions, object states, and first/third-person vision inputs, \etc. Predicting intention is made possible in VRGym as our unique multi-sensor setup reflects human poses, and the odometry data provided by the data logging system indicates human's trajectories.

\begin{figure}[t!]
  \centering
  \includegraphics[width=\linewidth]{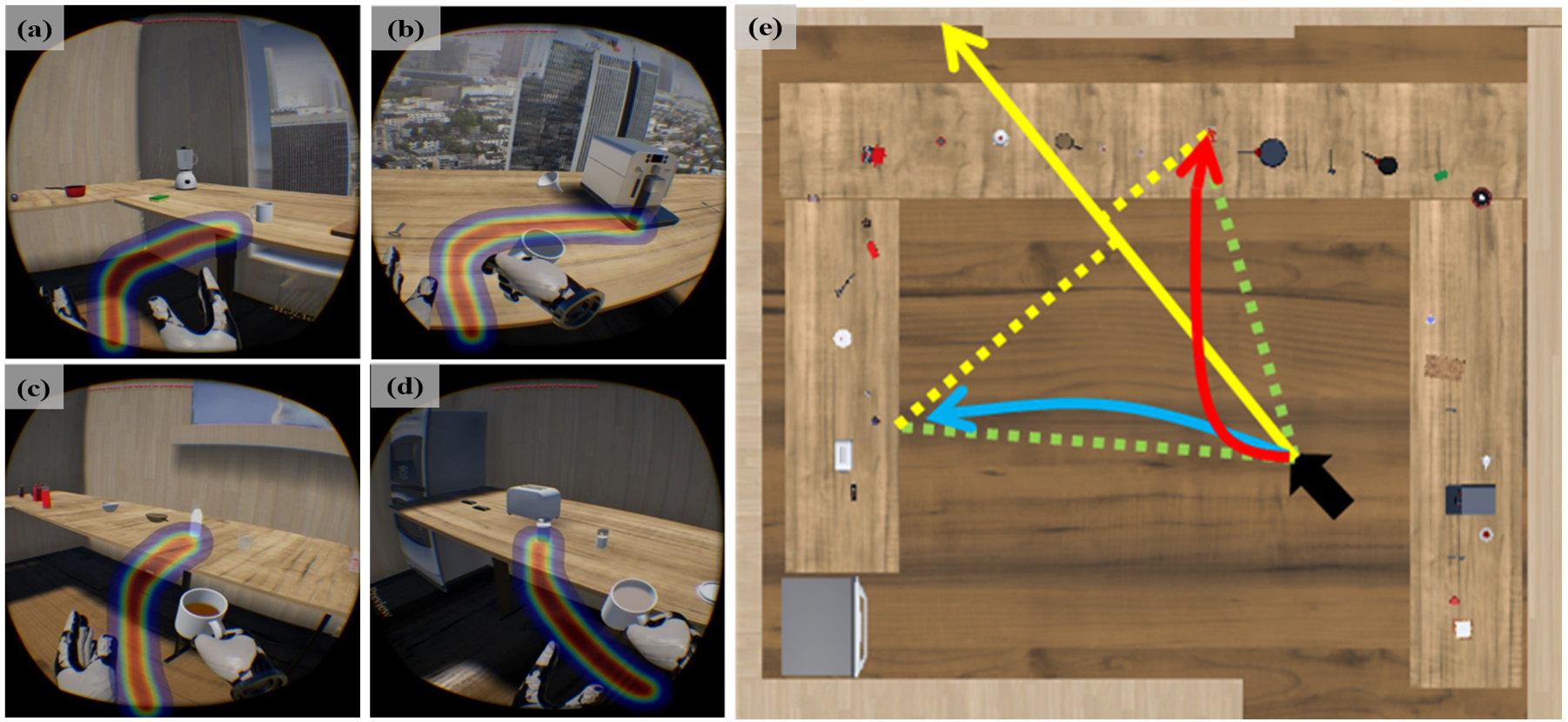}
  \caption{Intention predictions in a coffee-making task. (a) Grab a cup. (b) Use the coffee machine. (c) Pour milk. (d) Add sugar. (Right) Visualization of three intention prediction algorithms. Blue and Red: sampled paths from the grammar model~\cite{qi2017predicting}. Green: straight-line distance. Yellow: prediction by shortest perpendicular distance (dashed lines) from objects to the ray direction (solid arrow) based on avatar's location.}
  \label{fig:intention_results}
\end{figure}

In the experiment, we analyze different human intention prediction algorithms to demonstrate the potential of VRGym as a testbed for both physical and virtual \ac{ai} agents. Additionally, we show the unification of both the learning and the inference enabled by the VRGym. 20 subjects are recruited. The virtual environment is set up as a virtual kitchen, in which more than 20 objects are placed on top of three long tables. The layout of the kitchen is shown in \autoref{fig:intention_results}, where the agent starts from the entrance of the room (red dot) and performs the task with at least 4 steps: grasp a mug, operate the coffee maker, add milk, and add sugar. Note these tasks can perform in different orders. The resulting footprint from one subject is plotted in \autoref{fig:intention_results}. All subjects are required to perform a coffee-making task---making a cup of coffee using the available objects.

\autoref{fig:intention_results}e illustrates the comparisons among these three methods. The qualitative results are shown in \autoref{fig:intention_results}a-d to reveal the intention of the agent as the heat maps during the process of making coffee, where hotter color (red) indicates higher probability. This high-level semantic prediction is inferred given multiple human demonstrations as logged navigation and grasp data collected from the agent using VRGym. 


\subsection{Experiment 2: Social Interaction}\label{sec:social}

Social interactions or social HRI is a vital topic enabling human-robot co-existing environment, since the robot needs to understand and respond properly to human's social behaviors, such as waving and hand-shaking. Although the current robot simulators (\eg, Gazebo and V-Rep) provide a suite of features, one key element these simulation platforms still largely missing is the direct human involvement which is crucial for human-robot interaction studies.

\paragraph*{Participants}

A total of 10 subjects were recruited. We implemented the algorithm proposed in~\cite{shu2017learning} for robot learning social affordance. The algorithm is briefly described as follows; we refer the readers to the original paper for more technical details. 

\paragraph*{Results}

\begin{figure}[t!]
    \centering
    \includegraphics[width=\linewidth]{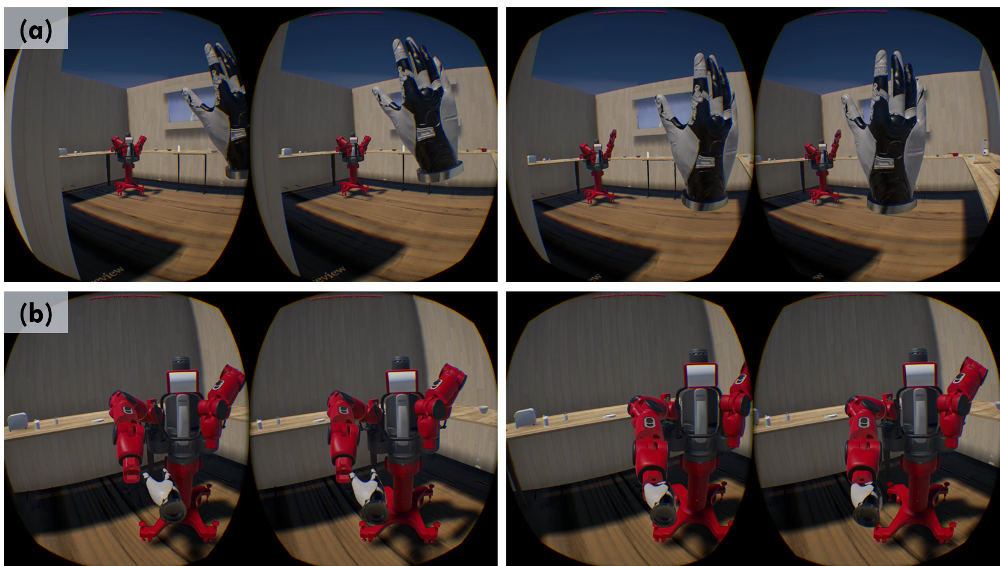}
    \caption{Human robot interactions in VRGym. A Baxter robot (a) waves hands and (b) shake hands with a virtual human agent.}
    \label{fig:hri}
\end{figure}

Qualitative results are shown in \autoref{fig:hri}. Concretely, the robot starts hand-waving in response to the agent's hand waving (\autoref{fig:hri}a), illustrated by a virtual hand model. The robot stretches out its manipulator to make a handshake with the virtual agent (\autoref{fig:hri}b). Technically, when the virtual Baxter inside the VRGym perceives the action signals from a virtual human such as \emph{hand shaking} or \emph{hand stretching out}, it sends the action signals to ROS through the VRGym-ROS bridge. In ROS, the motion planning will generate corresponding body parts transformations and send the computed transformation data back to the virtual Baxter inside VRGym, such that it will then act with the appropriate responses to the virtual human agent. In this sense, the proposed VRGym enables a new approach to study social human-robot interaction without using a costly physical robot or having a physical contact between a subject and robots, which in some cases could be dangerous.

\begin{figure}[b!]
    \centering
    \includegraphics[width=\linewidth]{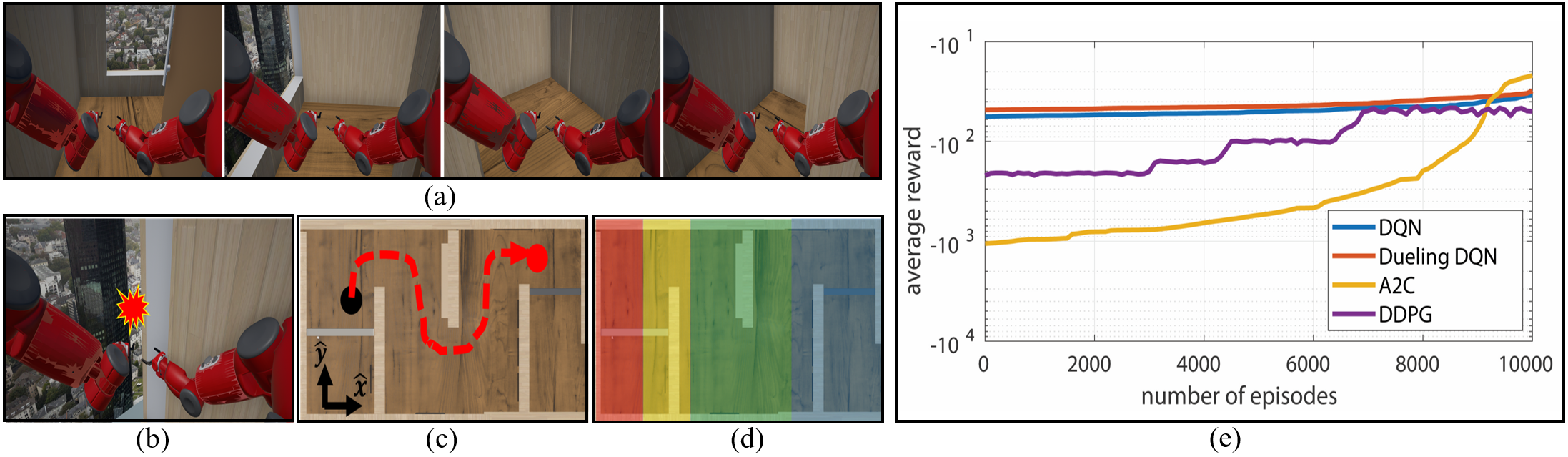}
    \caption{Settings for the \ac{rl} training inside VRGym environment for an indoor maze navigation task. (a) First-person view of a virtual robot. (b) The robot collides with a wall, triggering negative rewards. (c) An eagle view of the indoor navigation task. (d) Rewards assigned in different color zones (red, yellow, green and blue) from low to high. (e) The performances of the tested \ac{rl} algorithms.}
    \label{fig:rl_setting}
\end{figure}

\begin{figure}[t!]
    \centering
    \includegraphics[width=\linewidth]{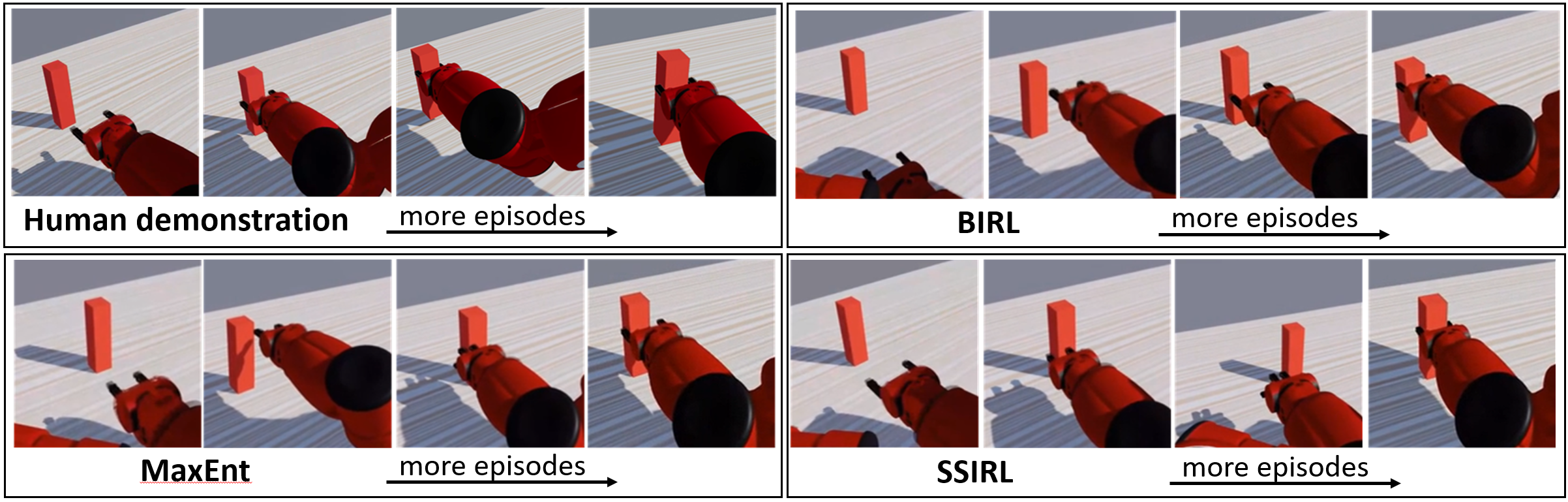}
    \caption{Learning human grasping demonstration with different \ac{irl} frameworks.}
    \label{fig:irl}
\end{figure}

\subsection{Experiment 3: \ac{rl} Algorithms Benchmark}\label{sec:rl}

We introduce a playground as a sub-module (\autoref{fig:rl_setting}) inside the VRGym, aiming to train robots to navigate in a 3D maze-like indoor corridor. The overall goal is to teach the robot agent itself by trial and error to obtain a navigation policy, reaching the final goal of the maze. The learning strategy applied on the virtual robot follows the standard \ac{rl} framework. A Baxter robot is integrated into the VRGym and controlled by off-the-shelf ROS packages.

Compared to other virtual playgrounds (\eg, OpenAI Gym), the proposed VRGym differs in two primary aspects.
\begin{itemize}[leftmargin=*,noitemsep,nolistsep]
    \item \emph{Sophisticated Interactions}. With the advanced physics-based simulator, the VRGym offers realistic interactions between the virtual agent and the virtual environment.
    \item \emph{Physical \ac{rl} Agent}. Since the VRGym is capable of importing both the physical and the virtual robot model to the virtual scene, it is feasible to transfer \ac{rl} model trained inside the virtual environment directly to a physical robot agent. 
\end{itemize}

We conduct four state-of-the-art deep \ac{rl} algorithms to demonstrate the VRGym's capability in \ac{rl} related tasks. These algorithms are \textbf{DDPG}~\cite{lillicrap2016continuous}, \textbf{DQN}~\cite{mnih2015human}, \textbf{Actor-Critic}~\cite{mnih2016asynchronous}, and \textbf{Dueling DQN}~\cite{wang2016dueling}. All four algorithms use the pixel-input from the first-person camera view. The quantitative comparison of the above four algorithms in VRGym is plotted in \autoref{fig:rl_setting}e. 

\subsection{Experiment 4: \ac{irl} for Learning Grasp}\label{sec:irl}


Grasp is an imperative capability for an interactive agent. In this experiment, we adopt an \acf{irl} framework to enable a virtual robot learning to grasp from human demonstrations. This task primarily involves both the data logging function in VRGym and a ROS motion planer communicated by the VRGym-ROS bridge. The robot is expected to learn how to successfully grasp an unknown object based on the hand trajectories demonstrated by the human subjects, collected through tele-operations using the Oculus Touch Controller inside the VRGym.

The trajectories of the human demonstrations are logged and used to infer the model and its parameters. Later, with the learned model and its parameters, the robot can be executed using the motion planner in ROS to grasp an unknown objects in the virtual environment.

Three \ac{irl} algorithms are implemented in the VRGym: Bayesian-IRL~\cite{ramachandran2007bayesian}, Maximum Entropy-IRL~\cite{ziebart2008maximum}, and Semi-supervised-IRL~\cite{valko2012semi}. Qualitative results are shown in \autoref{fig:irl}.

\section{Conclusion}\label{sec:conclusion}

In this paper, we introduce the VRGym as a promising simulation platform for training and evaluating autonomous agents to build the physical and interactive \ac{ai}. VRGym can represent a fine-grained human embodiment as a virtual avatar using a range of hardware setups for body and manipulation sensing. Existing robotics systems and algorithms developed in ROS can also be integrated to VRGym through a customized VRGym-ROS bridge. Multiple evaluations indicate that the VRGym has a robust performance at the system level and in the communication with ROS. Our experiments have demonstrated that four different robotics interactive tasks can be successfully trained using \ac{rl} and \ac{irl} inside VRGym. Specifically, we showcase how the data logged from the VRGym is useful in several interaction tasks, combining with the functions (\eg, motion planners, robot models) provided by ROS through the VRGym-ROS bridge. The successful implements of \ac{rl} and \ac{irl} for robotics interactive tasks in VRGym also support the training capability offered by VRGym in training robots with advanced machine learning methods. We believe VRGym could have further potential applications and it will benefit future research on the physical and interactive \ac{ai}.

{\small \noindent\textbf{Acknowledgement:}
The authors thank Xiaofeng Gao and Shu Wang of UCLA Statistics Department for useful discussions and assistance. The work reported herein was supported by DARPA XAI grant N66001-17-2-4029, ONR MURI grant N00014-16-1-2007, and a NVIDIA GPU donation.}
\vspace{-6pt}
\renewcommand*{\bibfont}{\fontsize{6.15}{6.15}\selectfont}
\bibliographystyle{ACM-Reference-Format}
\bibliography{main}


\end{document}